\documentclass[a4paper,12pt,draftcls,onecolumn]{IEEEtran}

\usepackage{amssymb}
\usepackage{algorithmic}
\usepackage{algorithm}
\usepackage{verbatim}
\usepackage[mathcal]{euscript}
\usepackage{mathrsfs}
\usepackage{cite}
\usepackage{multirow}

\newtheorem{theorem}{Theorem}
\newtheorem{proposition}{Proposition}
\newtheorem{lemma}{Lemma}

\ifCLASSINFOpdf
\else

   \usepackage[dvips]{graphicx}
   \graphicspath{{figures/}}
   \DeclareGraphicsExtensions{.eps}
\fi

\usepackage{amsmath}\allowdisplaybreaks
\usepackage{amsmath}
\usepackage[subnum]{cases}
\usepackage[caption=false,font=footnotesize]{subfig}

\allowdisplaybreaks[0]

\begin{document}

\title{Diversity Multiplexing Tradeoff of the Half-duplex Slow Fading Multiple Access Channel based on Generalized Quantize-and-Forward Scheme}

\author{Ming Lei and Mohammad Reza Soleymani\\
Electrical and Computer Engineering, Concordia University, Montreal, Quebec, Canada\\
Email:m\_lei,msoleyma@ece.concordia.ca
}

\maketitle

\begin{abstract}
This paper investigates the Diversity Multiplexing Tradeoff (DMT) of the generalized quantize-and-forward (GQF) relaying scheme over the slow fading half-duplex multiple-access relay channel (HD-MARC). The compress-and-forward (CF) scheme has been shown to achieve the optimal DMT when the channel state information (CSI) of the relay-destination link is available at the relay. However, having the CSI of relay-destination link at relay is not always possible due to the practical considerations of the wireless system. In contrast, in this work, the DMT of the GQF scheme is derived without relay-destination link CSI at the relay. It is shown that even without knowledge of relay-destination CSI, the GQF scheme achieves the same DMT, achievable by CF scheme with full knowledge of CSI.
\end{abstract}

\section{Introduction}

In wireless communication systems, relaying can either increase the system throughput or the reliability by creating a virtual distributed antenna system  \cite{Laneman2003,Azarian2005}. In the case of relay cooperating with multiple sources, a Multiple Access Channel with a relay (MARC) has been extensively studied in \cite{Kramer2005,Gunduz2010}.

\begin{figure}[t]
\centering
\subfloat[MARC Slot-1]           {\label{fig:MARC-phase1}
\includegraphics[width=1.5in]{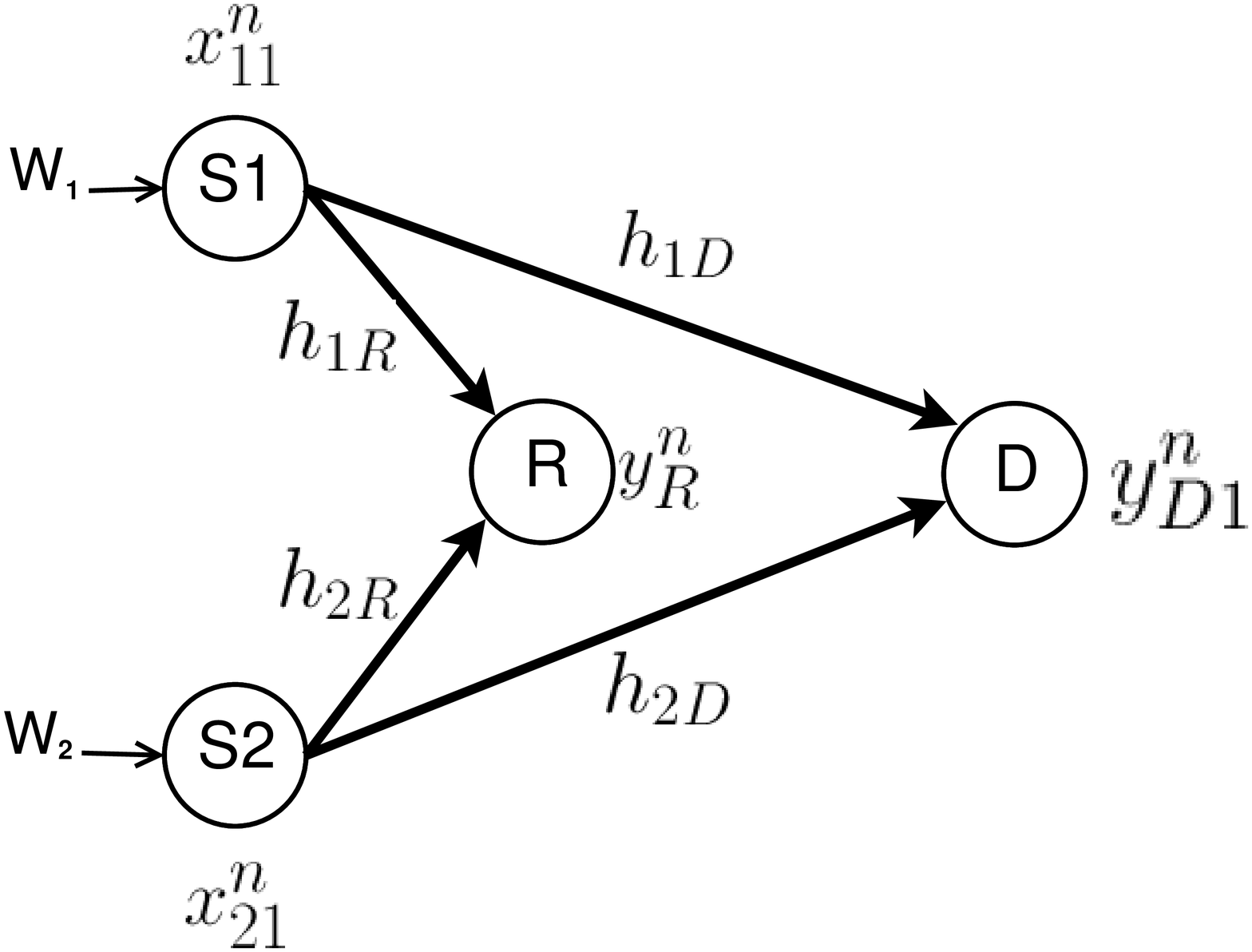}   }
\qquad
\subfloat[MARC Slot-2]{\label{fig:MARC-phase2}
\includegraphics[width=1.5in]{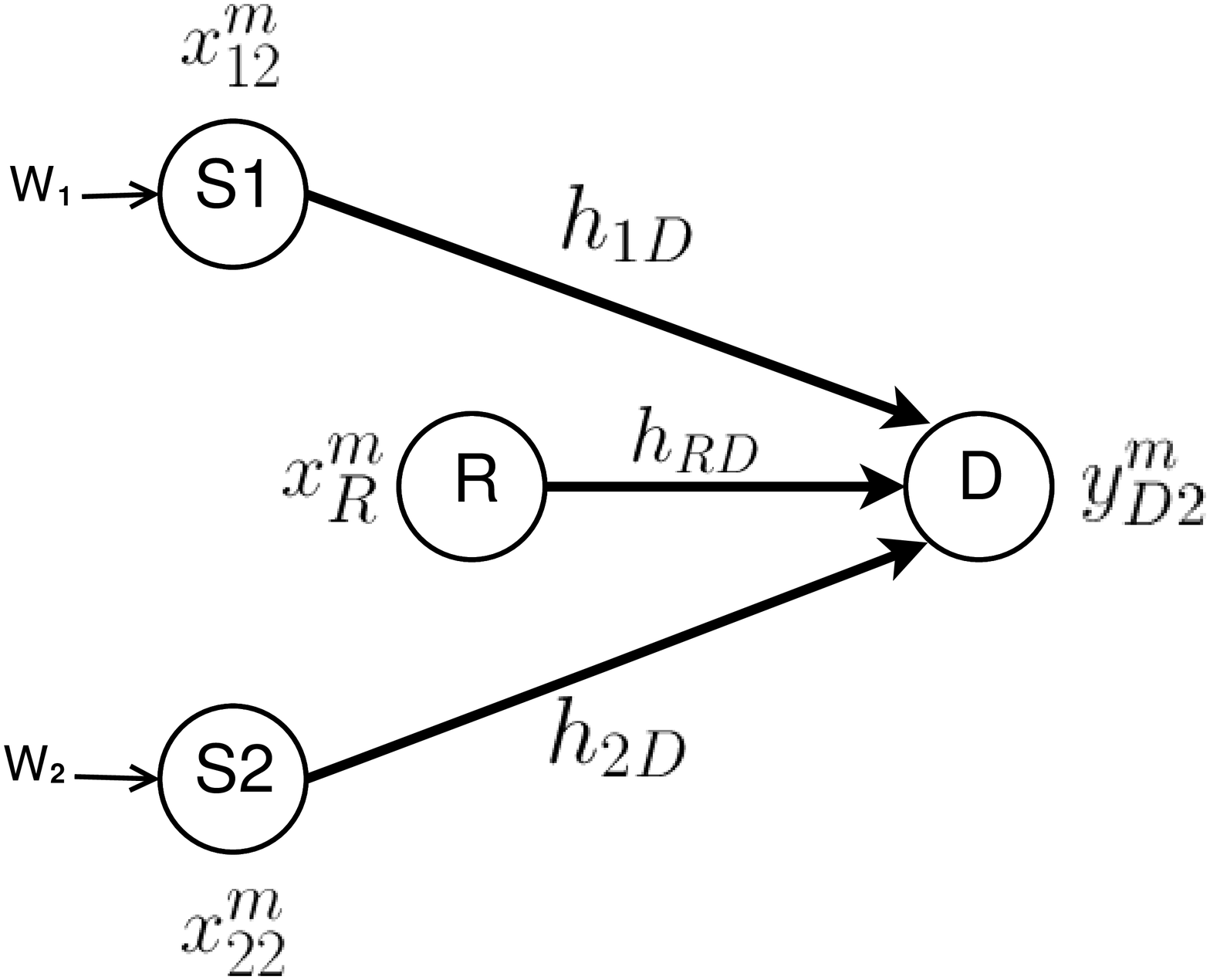}   }
\caption{Message flow of the Half-Duplex Multiple Access Relay Channel.}
\label{fig:MARC-phases}
\end{figure}

Motivated by the practical constraint that delay constrains exists in some wireless channels and relay cannot transmit and receive simultaneously in wireless communications\cite{Laneman2003,Khojastepour2003}, a slow fading Half-Duplex MARC (HD-MARC) (shown in Fig. \ref{fig:MARC-phases}) is considered in this work. In particular, a block fading channel where the channel coefficients stay constant in each block but change independently from block to block is studied. In addition, it is assumed that the channel state information (CSI) is not available at the transmitter side. Specifically, the destination has the receiver CSI and the relay has only the CSI of the source-to-relay link. The performance measure used in this work is the diversity-multiplexing tradeoff (DMT) which was introduced in \cite{Zheng2003}. The DMT characterizes the multiple-antena communications in terms of the relationship between system throughput and transmission reliability at asymptotically high signal-to-noise ratio (SNR).

For the HD-MARC, the DMT of different relaying schemes, i.e. dynamic decode-and-forward (DDF), multiaccess amplify-and-forward (MAF) and compress-and-forward (CF), have been characterized in \cite{Azarian2005, Chen2008,Yuksel2007}. In \cite{Yuksel2007}, it is shown that the CF scheme has its advantages over DDF and MAF scheme in terms of sustaining to multiple antennas case. Besides, the CF scheme is also able to achieve the optimal DMT upper bound when the multiplexing gain is higher than $\frac{4}{5}$. To achieve the optimal DMT, the CF scheme needs to have two assumptions: 1) using Wyner-Ziv coding and 2) the relay has perfect channel state information (CSI). The effect of the Wyner-Ziv coding on the DMT of the CF scheme has been investigated in \cite{Kim2009}. In practice, having the CSI of relay-destination link at relay is generally too ideal. When the critical delay constraint exist in the wireless channels, the relay may not be able to obtain the CSI accurately.

In this letter, we investigate the DMT of the HD-MARC based on the GQF scheme. The GQF scheme generalizes the quantize-and-forward (QF) scheme (a variation of the classic CF scheme) to the multiple user channels by taking into account the multiple access at both relay and destination. The QF scheme achieves the optimal DMT for a half-duplex three-node relay channel without the relay-destination link state available at relay \cite{Yao2013}. The CF scheme can achieve the optimal DMT of the symmetric HD-MARC when perfect CSI available at relay and $\frac{4}{5}<r<1$ \cite{Yuksel2007}. As shown in this work, the DMT achieved by the GQF scheme is 
\begin{eqnarray}
d_{GQF}(\bar{r})=
\begin{cases}
2-r, & \text{if } 0 \leq r \leq \frac{1}{2}  \\
3(1-r), & \text{if } \frac{1}{2} \leq r \leq 1.
\end {cases}
\nonumber
\end{eqnarray}
With only the source-relay CSI at relay, the GQF scheme achieves the optimal DMT for all the range of multiplexing gain $0<r<1$.

\section{System Model and Notations}

A half-duplex multiple access relay channel is considered in this work (Fig. \ref{fig:MARC-phases}). In particular, two sources $S_{1}$ and $S_{2}$ wish to communicate with one destination $D$. Relay helps the information propagation by cooperating with the two sources. Relay operates in the Half-Duplex mode, either receiving signals from the source nodes ($S_{1}$ and $S_{2}$) or transmitting to the destination $D$. Assume that each communication block length is totally $l$ channel uses and divided into two slots. The lengths of the first and the second slot are $n$ and $m$ channel uses, respectively. In the first slot, both $S_1$ and $S_2$ broadcast their messages to relay and $D$. In the second slot, $S_1$ and $S_2$ keep transmitting to $D$ while relay cooperatively transmitting to $D$ as well. Denote $x_{i1}^n$ and $x_{i2}^m$, $i \in \{1,2\}$ as the transmitted sequences by $S_i$ in the first and second slot correspondingly, and $x_{R}^m$ as the transmitted sequence by the relay node in the second slot, where $x_{ij}^k=[x_{ij,1},x_{ij,2},\cdots ,x_{ij,k}]$ and $x_{R}^k=[x_{R,1},x_{R,2},\cdots ,x_{R,k}]$. The received sequences at the destination in the first and the second slots are denoted as $y_{D1}^n$ and $y_{D2}^m$, respectively, and the received sequence at the relay in the first slot is denoted as $y_{R}^n$.
The channel transition probabilities are described by the following:
\begin{eqnarray}
y_{D1}^n & = & h_{1D}x_{11}^n+h_{2D}x_{21}^n+z_{D1}^n
\nonumber \\
y_{R}^n & = & h_{1R}x_{11}^n+h_{2R}x_{21}^n+z_{R}^n
\nonumber \\
y_{D2}^m & = & h_{1D}x_{12}^m+h_{2D}x_{22}^m+h_{RD}x_R^m+z_{D2}^m
\nonumber
\end{eqnarray}
where $h_{ij}$ for $i\in\{1,2,R\}$ and $j\in\{R,D\}$ denote the channel coefficients between the transmission node $i$ and the reception node $j$. For the slow fading channel, these coefficients are random variables and stay constant within each block and changes independently over different blocks. In particularly, a Rayleigh fading model is considered, which means the channel coefficients $h_{ij}$ are assumed to be mutually independent and circularly symmetric complex Gaussian with zero means and variances $\sigma_{ij}^2$. The elements of the noise sequences of $z_{11}^n, z_{12}^m$ and $z_{R}^n$ are also circularly symmetric complex Gaussian with zero means and unit variances. For the continuous-valued channels, the transmitters have power constraints over the transmitted sequences as the
$\frac{1}{n}\sum_{i=1}^{n}|x_{j,i}|^2  \leq  \text{SNR}, \text{for}\:j\in\{11,21\}$ and $\frac{1}{m}\sum_{i=1}^{m}|x_{k,i}|^2 \leq \text{SNR},\text{for}\:k\in\{12,22,R\}$, where $|x|$ shows the absolute value of $x$.

The following random variables are defined to clarify the input and output relationships of the HD-MARC.
Let $X_i$ for $i\in\{11,21,12,22,R\}$, $Z_j$ for $j\in\{D1,D2,R\}$ and $Z_Q$ be generic random variables which are complex Gaussian with zero mean and are mutually independent. The variances of $X_i$, $Z_j$ and $Z_Q$ are $P_i$, 1 and $\sigma_Q^2$ respectively. The random variable $Y_k$ denotes the channel output where $k\in\{D1,D2,R\}$.  The auxiliary random variable $\hat{Y}_R$ is the quantized signal of $Y_R$, i.e., $\hat{Y}_R = Y_R+Z_Q$ where $Z_Q$  is the quantization noise.

Follows the conventions as in \cite{Zheng2003,Azarian2005,Kim2009,Yao2013}, define $f(\text{SNR}) \doteq \text{SNR}^d$ if $\underset{\text{SNR} \to \infty}{\text{lim}} \frac{\log f(\text{SNR})}{\log \text{SNR}} = d $. As DMT discusses the system performance at asymptotically high SNR, we assume all transmitters has the power $P_i= \text{SNR}$. The information rate $R = r \text{ log SNR}$ is increasing with SNR by a fixed ratio $r$, where $0<r<1$. In a slow-fading environment, if the target data rate $R$ is greater than the instantaneous mutual information, then outage event occurs. Denote $P_{out}(R)$ as the outage probability. At high SNR, the outage exponent (diversity gain) is then defined as
$$d(r)= - \underset{\text{SNR} \to \infty}{\text{lim}} \frac{\log P_{out}(r\text{ logSNR})}{\log \text{SNR}},$$ where $r$ is refered as the multiplexing gain. A coding scheme achieves a diversity gain or outage exponent of $d(r)$ for any fixed $r$ when
$P_{out}(r \text{logSNR}) \doteq \text{SNR}^{-d(r)}$.

\section{DMT of the GQF Scheme}
To derive the DMT of the GQF scheme in the HD-MARC, the achievable rate region and the corresponding outage event and the outage probability is shown first. The DMT result and its discussion is shown in the second part of this section.

\subsection{Achievable Rate Region and Outage Probability}
In GQF, relay quantizes its observation $Y_R$ to obtain $\hat{Y}_R$ after the first slot, and then sends the quantization index $u\in\mathcal{U}=\{1,2, \cdots, 2^{lR_U}\}$ in the second slot with $X_R$. Unlike the conventional CF, no Wyner-Ziv binning is applied. At the destination, decoding is also different in the sense that joint-decoding of the messages from both slots without explicitly decoding the quantization index is performed in GQF scheme. The following theorem shows the achievable rate regions:
\begin{theorem}\label{th-QFD-MARC}
The following rate regions are achievable over discrete memoryless HD-MARC based on the GQF scheme:
{\setlength\arraycolsep{0.1em}
\begin{eqnarray}
R_1 & < & \beta I(X_{11};Y_{D1},\hat{Y}_{R} | X_{21})
\nonumber \\
& & + (1-\beta)I(X_{12};Y_{D2} | X_{22},X_{R})\label{eqn-GQF-R1}
\\
R_1 + R_U & < & \beta[I(X_{11},\hat{Y}_{R};X_{21},Y_{D1})+I(X_{11};\hat{Y}_R)]
\nonumber\\
& & + (1-\beta)I(X_{12},X_{R};Y_{D2} | X_{22})
\label{eqn-GQF-R1U}\\
R_2 & < & \beta I(X_{21};Y_{D1},\hat{Y}_{R} | X_{11})
\nonumber \\
& & +  (1-\beta)I(X_{22};Y_{D2} | X_{12}, X_{R})
\label{eqn-GQF-R2}\\
R_2 + R_U & < & \beta[I(X_{21},\hat{Y}_{R};Y_{D1} | X_{11})+I(X_{21};\hat{Y}_R)]
\nonumber \\
& & + (1-\beta)I(X_{22},X_{R};Y_{D2} | X_{12})
\label{eqn-GQF-R2U}\\
R_1 + R_2 & < & \beta I(X_{11},X_{21};Y_{D1},\hat{Y}_{R})
\nonumber \\
& & + (1-\beta)I(X_{12},X_{22};Y_{D2} | X_{R})
\label{eqn-GQF-R1R2}\\
R_1 + R_2 + R_U & < & \beta [I(X_{11},X_{21},\hat{Y}_{R};Y_{D1})+I(X_{11},X_{21};\hat{Y}_{R})]
\nonumber \\
&  & + (1-\beta)[I(X_{12},X_{22},X_{R};Y_{D2})\label{eqn-GQF-R1R2RU},
\end{eqnarray}
}
where  $\beta=n/l$ is fixed and
\begin{equation}
    R_U > \beta I(Y_R,\hat{Y}_R),\label{eq-rfindu}
\end{equation}
for all input distributions $$p(x_{11})p(x_{21})p(x_{12})p(x_{22})p(x_R)p(\hat{y}_R|y_R).$$
\end{theorem}
\begin{IEEEproof}The detail of the proof is shown in Appendix A.
\end{IEEEproof}

For convenience of derivation, denote the channel coefficient vector in the slow-fading HD-MARC as $\mathbf{h}:=[h_{1D},h_{2D},h_{1R},h_{2D},h_{RD}]\label{eqn-channelVec}$.
Given $\mathbf{h}$, the instantaneous mutual information corresponding to the left hand of \eqref{eqn-GQF-R1}-\eqref{eqn-GQF-R1R2RU} are denoted as $I_{R_i}(\mathbf{h})$, where $i \in \{1,1u,2,2u,12,12u\}$.

As the transmitters have no access to the CSI, $S_1$ and $S_2$ can only use a fixed rate pair of $(R_1,R_2)$ to send information. The relay node has no CSI of the relay-destination link, therefore it is not able to adapt to the channel state $\mathbf{h}$ but can only assume a fixed rate of $R_U$ for its transmission. In order to do so, the relay selects the auxiliary random variable $\hat{Y}_R$ and chooses the variance of the $Z_Q$. Since
\begin{equation}
   R_U =\beta I(Y_R,\hat{Y}_R)=\beta \text{log} (1+\frac{1+|h_{1R}|^2P_{11}+|h_{2R}|^2P_{21}}{\sigma_Q^2}) \label{eqn-fading-GQF-Ru}
\end{equation}
and all the parameters in (\ref{eqn-fading-GQF-Ru}) are known at relay, it can choose any fixed $R_U$ successfully.

In the GQF scheme, the destination node employs the joint-decoding technique, thus the outage event happens when either one of the conditions (\ref{eqn-GQF-R1})-(\ref{eqn-GQF-R1R2RU}) not satisfied. The outage event can be defined as the sets of
\begin{equation}
\begin{IEEEeqnarraybox}
[\IEEEeqnarraystrutmode
\IEEEeqnarraystrutsizeadd{2pt}
{2pt}
][c]
{rcl}
&\mathcal{O}_{R_1} &:= \{ \mathbf{h}:R_1 > I_{R_1}(\mathbf{h})\}
\\
&\mathcal{O}_{R_{1u}} &:= \{\mathbf{h}:R_1+R_U > I_{R_{1u}}(\mathbf{h})\}
\\
&\mathcal{O}_{R_2} &:= \{\mathbf{h}:R_2 > I_{R_{2}}(\mathbf{h})\}
\\
&\mathcal{O}_{R_{2u}} &:= \{\mathbf{h}:R_2+R_U > I_{R_{2u}}(\mathbf{h})\}
\\
&\mathcal{O}_{R_{12}} &:= \{\mathbf{h}:R_1+R_2 > I_{R_{12}}(\mathbf{h})\}
\\
&\mathcal{O}_{R_{12u}} &:= \{\mathbf{h}:R_1+R_2+R_U > I_{R_{12u}}(\mathbf{h})\}
\end{IEEEeqnarraybox}
\end{equation}

As in (\ref{eqn-fading-GQF-Ru}), $R_U$ is chosen to satisfy (\ref{eq-rfindu}), the outage probability of the GQF scheme can be described as
\begin{multline}
P_{out}^{GQF}(R_1,R_2,R_U)=
Pr\{\mathcal{O}_{R_1} \cup \mathcal{O}_{R_{1u}} \cup \mathcal{O}_{R_2} \cup \mathcal{O}_{R_{2u}} \cup \mathcal{O}_{R_{12}} \cup \mathcal{O}_{R_{12u}}  \} .\label{eqn-out-common}
\end{multline}

\subsection{DMT of the GQF scheme}
Based on the achievable rates and the outage probability, the DMT of the GQF scheme is derived and discussed in this section.

The DMT upper bound of the symmetric MARC from \cite{Azarian2005} and\cite{Yuksel2007} is
\begin{equation}
d_{upper}(\bar{r})=
\begin{cases}
2-r, & \text{if } 0 \leq r \leq \frac{1}{2}  \\
3(1-r), & \text{if } \frac{1}{2} \leq r \leq 1,
\end {cases}
\end{equation}
when both sources taking the same multiplexing gain of $\frac{r}{2}$, $\bar{r} = (\frac{r}{2},\frac{r}{2})$. Since the cut-set upper bound results a lower bound on the outage probability, the DMT upper bound of the system can be derived accordingly.

With the previously obtained achievable rates and the outage probability of the GQF scheme, the achievable DMT in the following proposition:
\begin{proposition} For the HD-MARC, the GQF scheme achieves the DMT
\begin{eqnarray}
d_{GQF}(\bar{r})=
\begin{cases}
2-r, & \text{if } 0 \leq r \leq \frac{1}{2}  \\
3(1-r), & \text{if } \frac{1}{2} \leq r \leq 1.
\end {cases}
\end{eqnarray}
This $d_{GQF}(\bar{r})$ is optimal as it is equal to the upper bound of the DMT of the HD-MARC.
\label{prop-GQF}
\end{proposition}
\begin{IEEEproof}
The detail of the proof is shown in Appendix B.
\end{IEEEproof}

As a reference, the DMT achieved by the CF scheme is shown in the below:
\begin{eqnarray}
d_{CF}(\bar{r})=
\begin{cases}
2(1-r), & \text{if } 0 \leq r \leq \frac{2}{3}  \\
1-\frac{r}{2}, & \text{if } \frac{2}{3} \leq r \leq \frac{4}{5}. \\
3(1-r), & \text{if } \frac{4}{5} \leq r \leq 1.
\end {cases}\nonumber
\end{eqnarray}

Based on the range of $r$, we can compare the DMT results shown above into two different cases. 

First, in low multiplexing region $r\leq\frac{1}{2}$, the typical outage event is happened when only one of the sources is in outage. Using time sharing of the relay, the CF scheme can achieve the DMT of $2\times 1$ MISO system, which is the optimum case. However, without time sharing of the relay, the GQF scheme is also able to achieve the optimal DMT. The GQF scheme, similarly as DDF scheme, shows the advantage of DMT. 

Second, when $r\geq\frac{1}{2}$, the typical outage event is caused by both of the users are in outage. If $r\geq\frac{4}{5}$, the CF scheme performs better than the DDF scheme \cite{Yuksel2007}. The CF scheme achieves optimal DMT since it compresses both sources together which is more efficient in high data rates. At the same condition, the GQF scheme also achieves the optimal DMT as it is naturally a variation of the classic CF scheme.

In both cases, the CF scheme requires complete CSI available at relay to achieve the optimal DMT in some range of $r$. However, the GQF achieves the optimal DMT for all ranges of $r$ without having the CSI of relay-destination link at relay.

\section{Conclusion}

In this paper, the DMT of the GQF scheme has been derived in the slow fading half-duplex MARC. It is shown that the GQF scheme can achieve the optimal DMT when the relay has no access to the CSI of the relay-destination link while the classic CF scheme can only achieve some part of optimal DMT with complete CSI at relay.

\appendices

\section{Proof of Theorem \ref{th-QFD-MARC}}
Assume the source messages $W_{1}$ and $W_{2}$ are independent of each other. Each message $W_{i}$, $i\in\{1,2\}$, is uniformly distributed in its message set $\mathcal{W}_i = [1 : 2^{lR_i}]$.

\subsubsection{Codebook Generation}
Assume the joint pmf factors as
\begin{equation}
p(x_{11})p(x_{21})p(x_{12})p(x_{22})p(x_R)p(\hat{y}_R|y_R)
p(y_{D1},y_R|x_{11},x_{12})p(y_{D2}|x_{12},x_{22},x_R).
\end{equation}
Fix any input distribution $p(x_{11})p(x_{21})p(x_{12})p(x_{22})p(x_R)p(\hat{y}_R|y_R)$,
for $k=1,2$, randomly and independently generate
\begin{itemize}

    \item $2^{lR_k}$ codewords $x_{k1}^{n}(w_k)$, $w_k\in\mathcal{W}_k$, each according to $\prod _{i=1}^{n} p_{X_{k1}} (x_{k1,i}(w_k))$;
    \item $2^{lR_k}$ codewords $x_{k2}^{m}(w_k)$, $w_1\in\mathcal{W}_k$, each according to $\prod _{i=1}^{m} p_{X_{k2}} (x_{k2,i}(w_k))$;
    \item $2^{lR_U}$ codewords $x_{R}^{m}(u)$, $u\in\mathcal{U}=\{1,2,\dots 2^{lR_U}\}$, each according to $\prod _{i=1}^{m} p_{X_{R}} (x_{R,i}(u))$.
\end{itemize}
Calculate the marginal distribution
$$p(\hat{y}_R)=\sum_{x_{11}\in \mathcal{X_{11}} ,x_{21}\in \mathcal{X_{21}},y_{D1}\in \mathcal{Y_{21}},y_{R}\in \mathcal{Y_R}}p(\hat{y}_R|y_R)p(y_R,y_{D1}|x_{11},x_{21})p(x_{11})p(x_{21}),$$
randomly and independently generate $2^{lR_U}$ codewords $\hat{y}_{R}^{n}(u)$, each according to $\prod _{i=1}^{n} p_{\hat{Y}_{R}} (\hat{y}_{R,i}(u)).$

\subsubsection{Encoding}
To send message $w_i$, the source node $S_i$ transmits $x_{i1}^{n}(w_i)$ in the first slot and $x_{i2}^{m}(w_i)$ in the second slot, where $i\in\{1,2\}$. Let $\epsilon' \in (0,1)$ . After receiving $y_R^n$ at the end of the first slot, the relay tries to find a unique $u\in\mathcal{U}$ such that
\begin{equation}
(y_R^n,\hat{y}_R^n(u))\in \mathcal{T}_{\epsilon'}^n(Y_R,\hat{Y}_R)
\end{equation}
where $\mathcal{T}_{\epsilon}^n(Y_R,\hat{Y}_R)$ is the $\epsilon$-strongly typical set as defined in \cite{Lim2011}. If there are more than one such $u$, randomly choose one in $\mathcal{U}$. The relay then sends $x_R^m(u)$ in the second slot.

\subsubsection{Decoding}
The destination $D$ starts decoding the messages after the second slot transmission finishes. Let $\epsilon'<\epsilon<1$. Upon receiving in both slots, $D$ tries to find a unique pair of the messages $\hat{w}_1\in\mathcal{W}_1$ and $\hat{w}_2\in\mathcal{W}_2$ such that
\begin{eqnarray}
(x_{11}^n(\hat{w}_1),x_{21}^n(\hat{w}_2),y_{D1}^n,\hat{y}_R^n(u)) \in \mathcal{T}_\epsilon^n(X_{11},X_{21},Y_{D1},\hat{Y}_R)\nonumber\\
(x_{12}^m(\hat{w}_1),x_{22}^m(\hat{w}_2),x_R^m(u),y_{D2}^m) \in \mathcal{T}_\epsilon^m(X_{12},X_{22},X_R,Y_{D2})\nonumber
\end{eqnarray}
for some $u\in\mathcal{U}$.

\subsubsection{Probability of Error Analysis}
Let $W_i$ denote the message sent from source node $S_i, i \in \{1,2\}$. $U$ represents the message index chosen by the relay $R$.
Based on the symmetry of the codebook construction and the fact that the messages $W_i$  is chosen uniformly from $\mathcal{W}_i$, the probability of error averaged on $W_i$ and $U$ over all possible codebooks is
\begin{equation}
Pr(\mathcal\epsilon) = Pr(\hat{W}_1\neq 1 \cup \hat{W}_2\neq 1 | W_1=1, W_2=1)\label{poe}.
\end{equation}

Define two events $\mathcal{E}_{0}$ and $\mathcal{E}_{(w_1,w_2)}$:
{\setlength\arraycolsep{0.1em}
\begin{align}
\mathcal{E}_{0}  &:=  \{((Y_R^n,\hat{Y}_R^n(u))\notin  \mathcal{T}_{\epsilon'}^n(Y_R\hat{Y}_R)), \text {for all} \: u \}
\\
\mathcal{E}_{(w_1,w_2)}    &:=
\{ (X_{11}^n(w_1),X_{21}^n(w_2),Y_{D1}^n,\hat{Y}_R^n(u))
\in \mathcal{T}_{\epsilon}^n(X_{11}X_{21}Y_{D1}\hat{Y}_R) \:\: \text{and}
\nonumber \\
&\qquad (X_{12}^m(w_1),X_{22}^m(w_2),X_R^m(u),Y_{D2}^m)
\in \mathcal{T}_{\epsilon}^m(X_{11}X_{21}X_RY_{D2}) \; \text{for some}\: u \}.
\end{align}
}
Then $Pr(\mathcal\epsilon)$ can be rewritten as
\begin{eqnarray}
Pr(\mathcal\epsilon)
& \leq & Pr (\mathcal{E}_{0}|W_1=1,W_2=1)
+ Pr( (\mathcal{E}_{(1,1)} )^c\cap\mathcal{E}_{0}^c|W_1=1,W_2=1)
\nonumber \\
&  & + Pr(\cup_{(w_1,w_2)\in\mathcal{A}} \mathcal{E}_{(w_1,w_2)}|W_1=1,W_2=1),
\label{ineqn-err}
\end{eqnarray}

where $\mathcal{A}:=\{(w_1,w_2)\in\mathcal{W}_1\times \mathcal{W}_2:(w_1,w_2)\neq (1,1)\}$. Assume $\beta$ is fixed, then by covering lemma \cite{Gamal2010}, $Pr(\mathcal{E}_{0}|W_1=1,W_2=1)\rightarrow 0$ when $l\rightarrow \infty$, if
\begin{equation}
    R_U > \beta I(Y_R,\hat{Y}_R) + \delta(\epsilon')
\end{equation}
where $\delta(\epsilon')\rightarrow 0$ as $\epsilon'\rightarrow 0$. By the conditional typicality lemma \cite{Gamal2010}, $Pr( (\mathcal{E}_{(1,1)} )^c\cap\mathcal{E}_{0}^c|W_1=1,W_2=1) \rightarrow 0$ as $l\rightarrow \infty$. Through some standard probability error analysis \cite{Yao2013}, the second line of (\ref{ineqn-err}),$Pr(\cup_{(w_1,w_2)\in\mathcal{A}} \mathcal{E}_{(w_1,w_2)}|W_1=1,W_2=1)\rightarrow 0$, for fixed $\beta = \frac{n}{l}$,  $1-\beta = \frac{m}{l}$, if $l\rightarrow\infty$, $\epsilon\rightarrow 0$ and the inequalities (\ref{eqn-GQF-R1})-(\ref{eqn-GQF-R1R2RU}) hold.
Therefore, the probability of error $P(\mathcal\epsilon) \rightarrow 0$. The proof completes and the achievable rate region is shown in \emph{Theorem \ref{th-QFD-MARC}}.

\section{Proof of Proposition \ref{prop-GQF}}
The lower bound of the DMT achieved by the GQF scheme will be derived first. Then we show that the lower bound meets the upper bound, hence the optimal DMT is achieved by the GQF scheme. In order to find the lower bound on DMT,  we need the following lemma:

\begin{lemma}
For the case $R_1=R_2=\frac{r}{2} \text{log SNR}$, $R_U=r_U \text{log SNR}$, and $\beta=r_U=\frac{1}{2}$,
{\setlength\arraycolsep{0.1em}
\begin{eqnarray}
\text{Pr}(\mathcal{O}_{R_{i}}) &\doteq &\text{SNR}^{-(2-r)}
\\
\text{Pr}(\mathcal{O}_{R_{12}}) &\doteq &\text{SNR}^{-4(1-r)}
\\
\text{Pr}(\mathcal{O}_{R_{12u}}) &\doteq& \text{SNR}^{-3(1-r)}
\end{eqnarray}
}
where $\mathcal{O}_{R_{i}}$, $i \in \{1,1u,2,2u \}$, $\mathcal{O}_{R_{12}}$and $\mathcal{O}_{R_{12u}}$ are the outage events defined previously.
\label{lemma-1}
\end{lemma}

\begin{IEEEproof} The detail is shown in the Appendix C.
\end{IEEEproof}

To find a lower bound on the DMT, the union upper bound is applied. The outage probability of the GQF scheme can be upper bounded by
{\setlength\arraycolsep{0.1em}
\begin{eqnarray}
Pr(\mathcal{O})&=&Pr( \mathcal{O}_{R_1} \cup \mathcal{O}_{R_{1u}} \cup \mathcal{O}_{R_2} \cup \mathcal{O}_{R_{2u}} \cup \mathcal{O}_{R_{12}} \cup \mathcal{O}_{R_{12u}} )
\nonumber \\
&\leq &Pr(\mathcal{O}_{R_1}) + Pr(\mathcal{O}_{R_{1u}})+ Pr(\mathcal{O}_{R_2}) + Pr(\mathcal{O}_{R_{2u}})
\nonumber \\
&& + Pr(\mathcal{O}_{R_{12}}) + Pr(\mathcal{O}_{R_{12u}}).
\end{eqnarray}
}
In the symmetric HD-MARC, with any fixed $(\bar r)=(\frac{r}{2}, \frac{r}{2})$, $\beta$, $r_u$ and $R_1=R_2=\frac{r}{2} \text{log SNR}$, $R_U=r_u \text{log SNR}$, the outage exponent of the GQF scheme and its lower bound are:
{\setlength\arraycolsep{0.1em}
\begin{eqnarray}
d_{GQF}(\bar{r},\beta, r_u)& =& - \underset{\text{SNR} \to \infty}{\text{lim}} \frac{\text{log Pr}(\mathcal{O})}{\text{log SNR}}
\nonumber \\
&\geq & - \underset{\text{SNR} \to \infty}{\text{lim}} \frac{\text{log} \sum_i \text{Pr}(\mathcal{O}_{R_i})}{\text{log SNR}}
\nonumber \\
&=& d_{GQF}^*(\bar{r},\beta, r_u)
\end{eqnarray}
}
where  $i \in \{1,1u,2,2u,12,12u \}$ and $d_{GQF}^*(\bar{r},\beta, r_u)$ denotes the lower bound. Let $d_{R_i}(\bar{r},\beta, r_u)$ represent the outage exponent achieved by the set $\mathcal{O}_{R_i}$, we have
{\setlength\arraycolsep{0.1em}
\begin{eqnarray}
d_{R_i}(\bar{r},\beta, r_u)& =& - \underset{\text{SNR} \to \infty}{\text{lim}} \frac{\text{log Pr}(\mathcal{O}_{R_i})}{\text{log SNR}}.
\end{eqnarray}
}
When $\text{SNR} \to \infty$, the union bound outage probability will be dominated by the term with smaller exponent. In other words, the upper bound of the outage probability is mostly determined by the term with smallest diversity order, which is shown as
\begin{equation}
d_{GQF}^*(\bar{r},\beta, r_u) = \min_{\mathcal{O}_{R_i}}  d_{R_i}(\bar{r},\beta, r_u).
\end{equation}
For  each multiplexing exponent $r$, the outage exponent can be further optimized with the $\beta$ and $r_u$
{\setlength\arraycolsep{0.1em}
\begin{eqnarray}
d_{GQF}(\bar{r}) & = & \max_{\beta, r_u} d_{GQF}(\bar{r},\beta, r_u)
\nonumber \\
& \geq & \max_{\beta, r_u}  d_{GQF}^*(\bar{r},\beta, r_u)
\nonumber \\
& \geq & d_{GQF}^*(\bar{r},\frac{1}{2}, \frac{1}{2}).
\end{eqnarray}
}
From \emph{Lemma \ref{lemma-1}}, the outage exponents achieved by each of the outage event are $d_{R_1}(\bar{r},\frac{1}{2}, \frac{1}{2})=d_{R_{1u}}(\bar{r},\frac{1}{2}, \frac{1}{2})=d_{R_2}(\bar{r},\frac{1}{2}, \frac{1}{2})=d_{R_{2u}}(\bar{r},\frac{1}{2}, \frac{1}{2})=2-r$, $d_{R_{12}}(\bar{r},\frac{1}{2}, \frac{1}{2})=4(1-r)$ and $d_{R_{12u}}(\bar{r},\frac{1}{2}, \frac{1}{2})=3(1-r)$. $d_{GQF}^*(\bar{r},\frac{1}{2}, \frac{1}{2})$ is taking the minimum of the above terms, thus
\begin{eqnarray}
d_{GQF}^*(\bar{r},\frac{1}{2}, \frac{1}{2})=
\begin{cases}
2-r, & \text{if } 0 \leq r \leq \frac{1}{2}  \\
3(1-r), & \text{if } \frac{1}{2} \leq r \leq 1.
\end {cases}
\end{eqnarray}
Notice that when $0<r<1$, $3(1-r)$ is always less than $4(1-r)$. Therefore, $d_{R_{12u}}(\bar{r},\frac{1}{2}, \frac{1}{2})$ is smaller than $d_{R_{12}}(\bar{r},\frac{1}{2}, \frac{1}{2})$ for all values of $r$.

Since $d_{GQF}^*(\bar{r},\frac{1}{2}, \frac{1}{2})$ coincides with the upper bound of the symmetric HD-MARC from \cite{Azarian2005,Yuksel2007}, the GQF scheme achieves the optimal DMT. This finishes the proof of \emph{Proposition \ref{prop-GQF}}.

\section{Proof of lemma \ref{lemma-1}}

Following the similar steps as in \cite{Yao2013,Kim2009,Azarian2005}, let $\alpha_j = -\text{log}|h_j|^2/\text{log SNR}$ for $j \in \{11,21,1R,2R,RD\}$, $R_1=R_2=\frac{r}{2} \text{log SNR}$, $R_U=r_U \text{log SNR}$, and $\beta=r_U=\frac{1}{2}$. For $i \in \{1,1u,2,2u,12,12u \}$, denote the outage probability
\begin{equation}
Pr(\mathcal{O}_{R_{i}}) \doteq \text{SNR}^{-d_{i}}.
\end{equation}
Then the outage exponent or the diversity order can be derived by \cite{Zheng2003,Kim2009,Yao2013}
\begin{equation}
d_{i}= \underset{\mathcal{O}_{R_{i}}^{+}}{\text{inf}} (\alpha_{11}+\alpha_{1R}+\alpha_{21}+\alpha_{2R}+\alpha_{RD})
\label{eqn-dsum}
\end{equation}
where $\mathcal{O}_{R_{i}}^{+}$ is the set
\begin{eqnarray}
\mathcal{O}_{R_{i}}^{+} = \{ (\alpha_{11}, \alpha_{21}, \alpha_{1R}, \alpha_{2R}, \alpha_{RD}) \in \mathbb{R}^{5+}:\mathcal{O}_{R_{i}} \; \text{occurs} \}.
\end{eqnarray}


\subsection{outage exponent of $d_1$}
First rewrite $\mathcal{O}_{R_{1}}$ as
\begin{eqnarray}
\mathcal{O}_{R_1} = : & \{ R_1 > \beta \; \text{log} (1+|h_{11}|^2 P_{11}+ \frac{|h_{1R}|^2 P_{11}}{1+\sigma_Q^2})
 +(1-\beta) \text{log}(1+|h_{11}|^2 P_{12}) \}.
\end{eqnarray}
Perform the change of variables accordingly, $\mathcal{O}_{R_{1}}^{+}$ can be obtained
{\setlength\arraycolsep{0.1em}
\begin{eqnarray}
\mathcal{O}_{R_{1}}^{+} = \{ (\alpha_{11}, \alpha_{21}, \alpha_{1R}, \alpha_{2R}, \alpha_{RD}) \in \mathbb{R}^{5+}:
 \frac{r}{2}>\frac{1}{2}(1-\alpha_{11}, 1-\alpha_{1R})^{+} + \frac{1}{2}(1-\alpha_{11})^{+} \}.
\end{eqnarray}
}
Second, in order to solve the optimization problem of $d_1$, the above set can be partitioned into two cases. $d_1$ takes the minimum of the two solutions.

\textit{Case 1}: $\alpha_{11} \geq 1$. The inequality in $\mathcal{O}_{R_{1}}^{+} $ become
\begin{equation}
r > (1-\alpha_{11}, 1-\alpha_{1R})^{+}.
\label{eqn-case1}
\end{equation}
Based on the relationship between $\alpha_{11}$ and $\alpha_{1R}$, the above can be further divided into: 

\textit{Case 1.1}: $\alpha_{11} \leq \alpha_{1R}$. We have $\alpha_{1R} \geq 1$ and the optimum values of $\alpha$'s for this case, denoted as a vector $\boldsymbol{\alpha}^*$, are
\begin{equation}
\boldsymbol{\alpha}^*=(\alpha_{11}^*, \alpha_{21}^*, \alpha_{1R}^*, \alpha_{2R}^*, \alpha_{RD}^*) = (1,0,1,0,0).
\label{eqn-alpha1.1}
\end{equation}

\textit{Case 1.2}: $\alpha_{11} \geq \alpha_{1R}$. When $\alpha_{1R} \geq 1$, then the optimum $\boldsymbol{\alpha}^*$ are the same as \eqref{eqn-alpha1.1}. However, if $\alpha_{1R} \leq 1$, then \eqref{eqn-case1} become
\begin{equation}
r>1-\alpha_{1R}.
\end{equation}
The optimum $\boldsymbol{\alpha}^*$ is then
\begin{equation}
\boldsymbol{\alpha}^*= (1,0,(1-r)^+,0,0).
\label{eqn-alpha1.2}
\end{equation}
Let $d_{1-1}$ denote the minimum of the outage exponent in Case 1. Combining Case 1.1 and Case 1.2 gives
\begin{equation}
d_{1-1}=2-r
\end{equation}

\textit{Case 2}: $\alpha_{11} \leq 1$. The inequality in $\mathcal{O}_{R_{1}}^{+} $ changes to
\begin{equation}
r > (1-\alpha_{11}, 1-\alpha_{1R})^{+} + (1-\alpha_{11}).
\label{eqn-case2}
\end{equation}
Similarly as Case 1, Case 2 is also divided into two cases.

\textit{Case 2.1}: $\alpha_{11} \leq \alpha_{1R}$. Then \eqref{eqn-case2} becomes
\begin{equation}
r > (1-\alpha_{11}) + (1-\alpha_{11}).
\label{eqn-case2.1}
\end{equation}
This leads the optimum $\boldsymbol{\alpha}^*$ to be $(1-\frac{r}{2},0,1-\frac{r}{2},0,0)$.

\textit{Case 2.2}: $\alpha_{11} \geq \alpha_{1R}$. \eqref{eqn-case2} changes to
\begin{equation}
r > (1-\alpha_{1R}) + (1-\alpha_{11}).
\label{eqn-case2.2}
\end{equation}
This implies
\begin{equation}
\alpha_{11}+\alpha_{1R} > 2-r.
\end{equation}
Choosing $\alpha_{21}, \alpha_{2R}$ and $\alpha_{RD}$ equal to zero, the minimum of the outage exponent for Case 2.2 is $2-r$. Combining Case 2.1 and Case 2.2, we have
\begin{equation}
d_{1-2}=2-r,
\end{equation}
where $d_{1-2}$ denotes the minimum of the outage exponent in Case 2.

In the last, combing Case 1 and Case 2 and given $d_1= min (d_{1-1}, d_{1-2})$ we conclude
\begin{equation}
d_1=2-r=2(1-\frac{r}{2}).
\end{equation}

\subsection{outage exponent of $d_{1u}$, $d_2$ and $d_{2u}$}
Similarly as deriving $d_1$, rewrite $\mathcal{O}_{R_{1u}}$ as
{\setlength\arraycolsep{0.1em}
\begin{eqnarray}
\mathcal{O}_{R_{1u}} = : \{ R_1 +R_u &>& \beta \; \text{log}  [(1+|h_{11}|^2 P_{11})( \frac{1+\sigma_Q^2+|h_{1R}|^2 P_{11}+|h_{2R}|^2 P_{21}}{1+\sigma_Q^2})]
 \nonumber \\
 &&+(1-\beta) \text{log}(1+|h_{11}|^2 P_{12}+|h_{RD}|^2 P_R) \}.
\end{eqnarray}
}
Then we have $\mathcal{O}_{R_{1u}}^{+}$ as 
{\setlength\arraycolsep{0.1em}
\begin{eqnarray}
\mathcal{O}_{R_{1u}}^{+} &=& \{ (\alpha_{11}, \alpha_{21}, \alpha_{1R}, \alpha_{2R}, \alpha_{RD}) \in \mathbb{R}^{5+}:
\nonumber \\
 && r+1>(1-\alpha_{11})^{+}+(1-\alpha_{1R}, 1-\alpha_{2R})^{+} + (1-\alpha_{11},1-\alpha_{RD})^{+} \}.
\end{eqnarray}
}

Next, we solve the optimization problem of $d_{1u}$. Notice that in $\mathcal{O}_{R_{1u}}^{+}$, $(1-\alpha_{1R}, 1-\alpha_{2R})^{+}$ has three possible outcomes $1-\alpha_{1R}$, $1-\alpha_{2R}$ and $0$. Each of $(1-\alpha_{11})^{+}$  and $(1-\alpha_{11},1-\alpha_{RD})^{+}$ has two possible outcomes. Based on these outcomes, $\mathcal{O}_{R_{1u}}^{+}$ can be partitioned into twelve cases. The derivation of the outage exponent for each of these cases is similar to previous subsection. The result of these cases are shown in the Table \ref{table-d1u}. 

\begin{table} 
\renewcommand{\arraystretch}{1.3} 
\caption{Different Cases of Optimization for $d_{1u}$} 
\label{table-d1u} 
\centering 
\begin{tabular}{|c|c|c|c|c|} 
\hline 
\multirow{2}{*}{\bfseries Case No.} & \multicolumn{3}{c|}{\bfseries Outcomes from} & \multirow{2}{*}{\bfseries Minimum outage exponent}\\
\cline{2-4} 
& $(1-\alpha_{1R}, 1-\alpha_{2R})^{+}$ &$(1-\alpha_{11})^{+}$ &$(1-\alpha_{11},1-\alpha_{RD})^{+}$& \\
\hline\hline 
Case 1-1-1 & $1-\alpha_{1R}$ & $0$                    & $1-\alpha_{RD}$ & $2-r$\\  \hline 
Case 1-1-2 & $1-\alpha_{1R}$ & $0$                    & $0$                     & $2-r$\\  \hline 
Case 1-2-1 & $1-\alpha_{1R}$ & $1-\alpha_{11}$ & $1-\alpha_{11}$  & $2-r$\\  \hline 
Case 1-2-2 & $1-\alpha_{1R}$ & $1-\alpha_{11}$ & $1-\alpha_{RD}$ & $2-r$\\  \hline 

Case 2-1-1 & $1-\alpha_{2R}$ & $0$                    & $1-\alpha_{RD}$ & $2-r$\\  \hline 
Case 2-1-2 & $1-\alpha_{2R}$ & $0$                    & $0$                      & $2$\\  \hline 
Case 2-1-1 & $1-\alpha_{2R}$ & $1-\alpha_{11}$ & $1-\alpha_{11}$  & $2-r$\\  \hline 
Case 2-2-2 & $1-\alpha_{2R}$ & $1-\alpha_{11}$ & $1-\alpha_{RD}$ & $2-r$\\  \hline 

Case 3-1-1 & $0$                     & $0$                    & $1-\alpha_{RD}$ & $3$\\  \hline 
Case 3-1-2 & $0$                     & $0$                    & $0$                      & $4$\\  \hline 
Case 3-2-1 & $0$                     & $1-\alpha_{11}$ & $1-\alpha_{RD}$ & $3-r$\\  \hline 
Case 3-2-2 & $0$                     & $1-\alpha_{11}$ & $1-\alpha_{11} $ & $3-r$\\  \hline 
\end{tabular} 
\end{table} 

The eventual outage exponent of $d_{1u}$ takes the smallest value from the last column of the Table  \ref{table-d1u}. Therefore, we have
\begin{equation}
d_{1u}=2-r=2(1-\frac{r}{2}).
\end{equation}

Similarly as $d_{1}$ and $d_{1u}$, we can find $d_2$ and $d_{2u}$ as
\begin{equation}
d_{2}=d_{2u}=2-r=2(1-\frac{r}{2}).
\end{equation}

\subsection{outage exponent of $d_{12}$ and $d_{12u}$}
Following the similar process as previous subsections, we may rewrite $\mathcal{O}_{R_{12}}$ and $\mathcal{O}_{R_{12u}}$ as
{\setlength\arraycolsep{0.1em}
\begin{eqnarray}
\mathcal{O}_{R_{12}} = : \{ R_1 +R_2 &>& \beta \; \text{log} (1+|h_{11}|^2 P_{11} + |h_{21}|^2 P_{21}
\nonumber\\
&& +\frac{ (h_{11}h_{2R}-h_{1R}h_{21})^2 P_{11}P_{21}+|h_{1R}|^2 P_{11}+|h_{2R}|^2 P_{21}}{1+\sigma_Q^2})
 \nonumber \\
 &&+(1-\beta) \text{log}(1+|h_{11}|^2 P_{12}+|h_{21}|^2 P_{22}) \}.
 \\
 \mathcal{O}_{R_{12u}} = : \{ R_1 +R_2 +R_u&>& \beta \; \text{log} (\frac{ (1+|h_{11}|^2 P_{11} + |h_{21}|^2 P_{21})(1+\sigma_Q^2+|h_{1R}|^2 P_{11}+|h_{2R}|^2 P_{21})}{1+\sigma_Q^2})
 \nonumber \\
 &&+(1-\beta) \text{log}(1+|h_{11}|^2 P_{12}+|h_{21}|^2 P_{22}+|h_{RD}|^2 P_{R}) \}.
\end{eqnarray}
}
Then the corresponding  $\mathcal{O}_{R_{12}}^{+}$ and $\mathcal{O}_{R_{12u}}^{+}$ are
{\setlength\arraycolsep{0.1em}
\begin{eqnarray}
\mathcal{O}_{R_{12}}^{+} &=& \{ (\alpha_{11}, \alpha_{21}, \alpha_{1R}, \alpha_{2R}, \alpha_{RD}) \in \mathbb{R}^{5+}:
\nonumber \\
 && 2r>(1-\alpha_{11}, 1-\alpha_{21}, 1-\alpha_{1R}, 1-\alpha_{2R})^{+} + (1-\alpha_{11},1-\alpha_{21})^{+} \}.
\\
\mathcal{O}_{R_{12u}}^{+} &=& \{ (\alpha_{11}, \alpha_{21}, \alpha_{1R}, \alpha_{2R}, \alpha_{RD}) \in \mathbb{R}^{5+}:
\nonumber \\
 && 2r+1>(1-\alpha_{11},1-\alpha_{21})^{+}+(1-\alpha_{1R}, 1-\alpha_{2R})^{+} + (1-\alpha_{11},1-\alpha_{21},1-\alpha_{RD})^{+} \}
 \nonumber
 \\
\end{eqnarray}
}
Next, we solve the optimization problem of $d_{12}$ and $d_{12u}$. $\mathcal{O}_{R_{12}}^{+}$ and $\mathcal{O}_{R_{12u}}^{+}$ are partitioned into fifteen and thirty six cases respectively. The outage exponent results are shown in Table \ref{table-d12} and Table \ref{table-d12u}.

\begin{table} 
\renewcommand{\arraystretch}{1.3} 
\caption{Different Cases of Optimization for $d_{12}$} 
\label{table-d12} 
\centering 
\begin{tabular}{|c|c|c|c|c|} 
\hline 
\multirow{2}{*}{\bfseries Case No.} & \multicolumn{2}{c|}{\bfseries Outcomes from} & \multirow{2}{*}{\bfseries Minimum outage exponent}\\
\cline{2-3} 
&$(1-\alpha_{11},1-\alpha_{21})^{+}$ & $(1-\alpha_{11},1-\alpha_{21},1-\alpha_{1R}, 1-\alpha_{2R})^{+}$& \\
\hline\hline 
Case 1-1 & $1-\alpha_{11}$ & $1-\alpha_{11} $ &  $4(1-r)$\\  \hline 
Case 1-2 & $1-\alpha_{11}$ & $1-\alpha_{21}$  &  $4(1-r)$\\  \hline 
Case 1-3 & $1-\alpha_{11}$ & $1-\alpha_{1R}$ &  $4(1-r)$\\  \hline 
Case 1-4 & $1-\alpha_{11}$ & $1-\alpha_{2R}$ &  $4(1-r)$\\  \hline 
Case 1-5 & $1-\alpha_{11}$ & $0$                     &  $4$       \\  \hline 


Case 2    & \multicolumn{2}{c|}{ Similar to Case 1}                    &  $..$       \\  \hline 

Case 3-1 & $0$ & $1-\alpha_{11} $ &  $4$\\  \hline 
Case 3-2 & $0$ & $1-\alpha_{21}$  &  $4$\\  \hline 
Case 3-3 & $0$ & $1-\alpha_{1R}$ &  $4(1-r)$\\  \hline 
Case 3-4 & $0$ & $1-\alpha_{2R}$ &  $4(1-r)$\\  \hline 
Case 3-5 & $0$ & $0$                     &  $4$       \\  \hline 
\end{tabular} 
\end{table} 

\begin{table} 
\renewcommand{\arraystretch}{1.3} 
\caption{Different Cases of Optimization for $d_{12u}$} 
\label{table-d12u} 
\centering 
\begin{tabular}{|c|c|c|c|c|} 
\hline 
\multirow{2}{*}{\bfseries Case No.} & \multicolumn{3}{c|}{\bfseries Outcomes from} & \multirow{2}{*}{\bfseries Minimum outage exponent}\\
\cline{2-4} 
&$(1-\alpha_{11}, 1-\alpha_{21})^{+}$ & $(1-\alpha_{1R}, 1-\alpha_{2R})^{+}$  &$(1-\alpha_{11}, 1-\alpha_{21}, 1-\alpha_{RD})^{+}$& \\
\hline\hline 
Case 1-1-1 & $1-\alpha_{11}$ & $1-\alpha_{1R}$& $1-\alpha_{11}$  & $3-3r$\\  \hline 
Case 1-1-2 & $1-\alpha_{11}$ & $1-\alpha_{1R}$& $1-\alpha_{21}$  & $3-3r$\\  \hline 
Case 1-1-3 & $1-\alpha_{11}$ & $1-\alpha_{1R}$& $1-\alpha_{RD}$ & $3-3r$\\  \hline 
Case 1-1-4 & $1-\alpha_{11}$ & $1-\alpha_{1R}$ & $0$                     & $3$    \\  \hline 
Case 1-2    & \multicolumn{3}{c|}{ Similar to Case 1-1}                       &  $..$       \\  \hline 
Case 1-3-1 & $1-\alpha_{11}$ & $0$& $1-\alpha_{11}$  & $3.5-3r$\\  \hline 
Case 1-3-2 & $1-\alpha_{11}$ & $0$& $1-\alpha_{21}$  & $3.5-3r$\\  \hline 
Case 1-3-3 & $1-\alpha_{11}$ & $0$& $1-\alpha_{RD}$ & $3-2r$\\  \hline 
Case 1-3-4 & $1-\alpha_{11}$ & $0$& $0$                     &  $5$    \\  \hline 

Case 2    & \multicolumn{3}{c|}{ Similar to Case 1}                       &  $..$       \\  \hline 

Case 3-1-1 & $0$ & $1-\alpha_{1R}$ & $1-\alpha_{11}$  & $3$\\  \hline 
Case 3-1-2 & $0$  & $1-\alpha_{1R}$& $1-\alpha_{21}$  & $3$\\  \hline 
Case 3-1-3 & $0$  & $1-\alpha_{1R}$& $1-\alpha_{RD}$ & $3-2r$\\  \hline 
Case 3-1-4 & $0$  & $1-\alpha_{1R}$ & $0$                     & $3$    \\  \hline 
Case 3-2    & \multicolumn{3}{c|}{ Similar to Case 3-1}                       &  $..$       \\  \hline 
Case 3-3-1 & $0$  & $0$& $1-\alpha_{11}$  & $5$\\  \hline 
Case 3-3-2 & $0$  & $0$& $1-\alpha_{21}$  & $5$\\  \hline 
Case 3-3-3 & $0$  & $0$& $1-\alpha_{RD}$ & $4$\\  \hline 
Case 3-3-4 & $0$  & $0$& $0$                     &  $5$    \\  \hline 
\end{tabular} 
\end{table} 

The eventual outage exponent of $d_{12}$ and $d_{12u}$ takes the smallest value from the last column of the Table  \ref{table-d12} and Table  \ref{table-d12u} . Therefore, we have
{\setlength\arraycolsep{0.1em}
\begin{eqnarray}
d_{12}&=&4-4r=4(1-r)
\\
d_{12u}&=&3-3r=3(1-r).
\end{eqnarray}
}
The proof for Lemma \ref{lemma-1} is finished as we find $d_{1}=d_{1u}=d_{2}=d_{2u}=2(1-\frac{r}{2})$, $d_{12}=4(1-r)$ and $d_{12u}=3(1-r)$.

\bibliographystyle{IEEEtran}
\bibliography{reference-QFCF-DMT}

\end{document}